\documentclass[prc,twocolumn,superscriptaddress,showpacs,preprintnumbers,amsmath,amssymb]{revtex4-2}
\usepackage[dvips]{graphicx}
\usepackage{ulem,color}

\def\m@thcombine#1#2{%
  \setbox0=\hbox{$#1$}
  \setbox1=\hbox{$#2$}
  \ifdim\wd0>\wd1
    \setbox0=\hbox to\wd1{\hss\box0\hss}
  \else
    \setbox1=\hbox to\wd0{\hss\box1\hss}
  \fi
  \mathop{\vcenter{
    \offinterlineskip\box0\box1}}}
\def\lesim{\m@thcombine<\sim}
\def\gesim{\m@thcombine>\sim}

\def\vp{\mbox{\boldmath$p$}}

\def\vQ{\mbox{\boldmath$Q$}}
\def\vX{\mbox{\boldmath$X$}}

\def\-{\mbox{$-$}}

\begin{document}
\title{Uncertainty evaluation of peak energy of giant dipole resonance propagated from uncertainties of Skyrme parameters}
\author{Tsunenori Inakura}
\affiliation{Laboratory for Zero-Carbon Energy, Institute of Innovative Research, Tokyo Institute of Technology, Tokyo 152-8550, Japan}

\begin{abstract}
We evaluate uncertainty of peak energy of giant dipole resonance (GDR), propagated from uncertainty of parameters of Skyrme interaction.
The Monte Carlo calculation of the random phase approximation using randomized Skyrme parameters is performed.
Under the condition that the correlations between each of the Skyrme parameters is considered, 
the GDR peak energy has the uncertainty of $\sim$ 1 MeV irrespective of nuclear mass
and is strongly correlated with the Skyrme parameters, in present calculations.
This serves a guide for a new parametrization of effective interactions.
\end{abstract}

\maketitle

\section{Introduction}

Nuclear density functional theory (DFT) with an energy density functionals (EDFs) or effective interaction is 
one of the standard theoretical tools to investigate the structure of atomic nuclei, 
and has been applied to systematic calculations of nuclei covering wide range of 
the nuclear chart~\cite{Stoitsov03,Terasaki06,Terasaki07,Hilaire07,Terasaki08,Inakura11,Ebata17,Bulgac18}.
In many cases, the DFT calculations provides good descriptions of properties of the ground states and excited states 
in satisfactory accuracy.
However, it sometimes give values deviated from the experimental data.
It is known that the random phase approximation (RPA) calculation with Skyrme interaction underestimates 
peak energies of giant dipole resonances (GDRs) in light nuclei by a few MeV.

The parameters of effective interactions and EDFs are determined to reproduce limited sets of experimental data of ground states in spherical nuclei
and nuclear matter properties.
This raises the question of its predictive power especially for excited states and unstable nuclei.
Therefore, uncertainty estimation of the calculated values and feedback to the parameters are strongly desired.
In last decade, many theoretical studies \cite{Fattoyev11,Gao13,Kortelainen13,Reinhard13,Goriely14,Kortelainen15,Erler15,Roca-Maza15,Reinhard16,Haverinen17,Kejzlar20,Sprouse20} 
calculated correlations of the parameters (covariance analysis) and evaluated uncertainty of 
calculated values of observables and the nuclear matter properties, coming from the uncertainty of the parameters.
However, these results are not used for improvement of the parameters.
This is because relations of the parameters and the calculated values of observables are not known clearly.

We performed the Monte Carlo calculation of RPA with randomized Skyrme paremeters and
evaluated uncertainty of GDR peak energy, propagated from uncertainty of Skyrme parameters.
By explicitly taking the correlations of the parameters in randomization, we found that 
there are strong correlations between the Skyrme parameters and the calculated GDR peak energy.

This paper is organized as follows. In section \ref{sec:metohd} we briefly review the theoretical framework 
related to the topics and the Monte Carlo calculation. In section \ref{sec:result} we present out results and conclusions are given in section \ref{sec:conclusion}.

\section{Method}
\label{sec:metohd}

A detailed explanation of uncertainty analysis is given in Ref.~\cite{Dobaczewski14}.
In this section, we repeat briefly the basic formula and outline the uncertainties 
of the Skyrme parameters, their correlations (covariance), and 
procedure of Monte Carlo calculation for the uncertainty evaluation.

\subsection{Uncertainty and correlation of parameters}

The calculation model space is determined by $m$ parameters, $\vp=\left(p_1,p_2, \cdots, p_m\right)^T$.
These parameters define an effective interaction.
Therefore, observables $\mathcal{O}$ are functional of the parameters, $\mathcal{O}(\vp)$.
Because the number of parameters is much smaller than the number of observables, 
there is correlations between the computed quantities. 
Moreover, there exist correlations between the parameters 
because the model parameters are optimized to a limited number of observables.

Optimal model parametrization $\vp_0$ is calibrated by a least-squares fit to experimental and semi-empirical data,
\begin{eqnarray}
\chi^2(\vp) = \sum^K_{k=1} \left(\frac{\mathcal{O}^\mathrm{th}_k(\vp) - \mathcal{O}^\mathrm{ref}_k}{\Delta \mathcal{O}^\mathrm{ref}_k}\right)^2 \,,
\end{eqnarray}
where $\mathcal{O}^\mathrm{th}(\vp)$ are calculated values by utilizing the parameter set $\vp$, 
$\mathcal{O}^\mathrm{ref}$ are related to experimental data and/or semi-empirical data such as the nuclear equation-of-state parameters, and 
$\Delta \mathcal{O}^\mathrm{ref}$ stands for the adopted errors. 
The optimal parametrization $\vp_0$ is the one that minimizes $\chi^2$ with the minimum value given by $\chi^2_0 \equiv \chi^2(\vp_0)$.

Model parameter $\vp$ which lies in the vicinity $\vp_0$ provides a good description of the experimental data. 
The range of reasonable parametrization is defined to cover all model parameter $\vp$ for which 
$\chi^2(\vp) \le \chi^2_0 +1$ \cite{Brandt}. 
We can expand $\chi^2$ in a power series around $\vp_0$. That is, up to second order in $(\vp-\vp_0)$, we obtain
\begin{eqnarray}
&& \chi^2(\vp)  \approx  \chi^2_0 + \sum^n_{i,j} \left( \vp - \vp_0 \right)_i \mathcal{M}_{ij}\left( \vp - \vp_0 \right)_j \,,
\end{eqnarray}
where $\mathcal{M}_{ij}$ is the matrix of second derivative 
\begin{eqnarray}
\mathcal{M}_{ij} = \left. \frac{1}{2} \partial_{p_i} \partial_{p_j} \chi^2(\vp)\right|_{\vp=\vp_0}\,.
\end{eqnarray}
The reasonable parametrization fill the confidence ellipsoid given by 
\begin{eqnarray}
\left( \vp - \vp_0 \right)^T \mathcal{M} \left( \vp - \vp_0 \right) \le 1 \,.
\label{confidenceElipsoid}
\end{eqnarray}
This confidence ellipsoid leads to estimation of the uncertainty (statistical error) $\Delta p$ of the optimal parametrization, as
\begin{eqnarray}
\Delta p_i = \sqrt{\left(\mathcal{M}^{-1}\right)_{ii}} \equiv \sqrt{\mathcal{S}_{ii}} \,,
\end{eqnarray}
since the equation $\chi^2(\vp+\Delta \vp) = \chi^2_0+1 $ determines the magnitude of $\Delta p_i$. 
Here, the covariance matrix $\mathcal{S}$ is defined as $\mathcal{M}^{-1}$.
In this paper we employ the SLy5-min parameter set \cite{SLy5-min}.
The SLy5-min parameters and its uncertainty $\Delta p_i$ are listed in Table~\ref{table:SLy5-min.Parameter}.

The correlation matrix $\mathcal{C}$ is estimated, from the covariance matrix $\mathcal{S}$, as
\begin{eqnarray}
\mathcal{C}_{ij} \equiv \frac{\mathcal{S}_{ij}}{\sqrt{\mathcal{S}_{ii}\mathcal{S}_{jj}}}
\end{eqnarray}
where $\mathcal{C}_{ij}$ takes values form -1 to 1. 
$\mathcal{C}_{ij} \approx 1$ indicates a strong correlation and -1 a strong 
anti-correlation between parameters $p_i$ and $p_j$. This indicates that $p_i$ (or $p_j$) is redundant and can be fixed 
during the fit by setting its value. On the contrary, $\mathcal{C}_{ij} \approx 0$ means 
that no correlation holds between parameters $p_i$ and $p_j$.
This clearly indicates that both parameters are needed for the description of the set of observables used for the fit.
Table~\ref{table:correlationMatrix} shows the correlation matrix of SLy5-min parameter.
The correlation matrix elements concerning $x_2$, $W_0$, $W^\prime_0$, and $\alpha$ are zero
because they are fixed in fitting procedure as shown in Table~\ref{table:SLy5-min.Parameter}.
Therefore, such matrix elements are not shown.

\begin{table}
\caption{Parameter name $\vp$, their optical value $\vp_0$, and uncertainty $\Delta \vp$ for SLy5-min parameter \cite{SLy5-min}.}
\begin{tabular}{c|cccl}
 $\vp$    &   $\vp_0$  & $\pm$ & $\Delta\vp$ & unit \\ \hline
 $t_0$    &  -2475.408 & $\pm$ & 149.455 & MeV fm$^3$ \\
 $t_1$    &    482.842 & $\pm$ & 58.537  & MeV fm$^5$ \\
 $t_2$    &   -559.374 & $\pm$ & 144.534 & MeV fm$^5$ \\
 $t_3$    &  13697.07  & $\pm$ & 1672.93 & MeV fm$^{3+3\alpha}$ \\
 $x_0$    &  0.741185  & $\pm$ & 0.189191 &  \\
 $x_1$    & -0.146374  & $\pm$ & 0.468173 &  \\
 $x_2$    &  -1        &       & fixed    &  \\
 $x_3$    &  1.162688  & $\pm$ & 0.340537 &  \\
 $W_0$    &  126       &       & fixed    & MeV fm$^5$ \\
 $W^\prime_0$    &  126       &       & fixed    & MeV fm$^5$ \\
 $\alpha$ &  1/6       &       & fixed    &  \\
\end{tabular}
\label{table:SLy5-min.Parameter}
\end{table}

\begin{table}
\caption{Correlation matrix elements of Skyrme SLy5-min parameters, $\mathcal{C}$, taken from Ref.~\cite{SLy5-min.correlation}.}
\begin{ruledtabular}
\begin{tabular}{c|rrrr|rrr}
 & $t_0$ & $t_1$ & $t_2$ & $t_3$ & $x_0$ & $x_1$ & $x_3$ \\ \hline
 $t_0$ &  1.0000 &  0.9837 &  0.9854 & -0.9997 & -0.6766 &  0.8110 & -0.6158 \\
 $t_1$ &  0.9837 &  1.0000 &  0.9575 & -0.9870 & -0.7066 &  0.8489 & -0.6553 \\
 $t_2$ &  0.9854 &  0.9575 &  1.0000 & -0.9863 & -0.6601 &  0.7843 & -0.5964 \\
 $t_3$ & -0.9997 & -0.9870 & -0.9863 &  1.0000 &  0.6798 & -0.8154 &  0.6197 \\ \hline
 $x_0$ & -0.6766 & -0.7066 & -0.6601 &  0.6798 &  1.0000 & -0.9327 &  0.9928 \\
 $x_1$ &  0.8110 &  0.8489 &  0.7843 & -0.8154 & -0.9327 &  1.0000 & -0.9311 \\
 $x_3$ & -0.6158 & -0.6553 & -0.5964 &  0.6197 &  0.9928 & -0.9311 &  1.0000 \\
\end{tabular}
\end{ruledtabular}
\label{table:correlationMatrix}
\end{table}

\subsection{Monte Carlo calculation and uncertainty propagation}

For Monte Carlo calculation using randomized SLy5-min parameter sets, 
we generate $n=1000$ random samples, $\vX^{(j)}$ ($j=1,2,\cdots,n$), which satisfy the correlation shown in Table~\ref{table:correlationMatrix}.
Here, we introduce a Normal distribution $\mathcal{N}(\mu, \sigma)$, where 
$\mu$ and $\sigma$ are the mean value and the standard deviation of the distribution, respectively.
The correlation matrix $\mathcal{C}$, 
which is positive semi-definite matrix,
is factorized as $\mathcal{C}=\vQ^T\vQ$ by the singular value decomposition.

First, we generate $n$ sets of random independent samples $\overline{\vX}^{(j)} =(\overline{x}^{(j)}_1, \overline{x}^{(j)}_2,\cdots,\overline{x}^{(j)}_m)^T$ 
with each variables $\sim \mathcal{N}(0,1)$.
Each variable in $\overline{\vX}^{(j)}$ is uncorrelated with each others.
Secondly, we act the factorized matrix $\vQ$ on each $\overline{\vX}^{(j)}$, i.e., creating the correlated samples $\left(\vQ\overline{\vX}\right)^{(j)}$ 
satisfying the correlation $\mathcal{C}$.
Thirdly, we shift the mean values and standard deviations, $X^{(j)}_i = p_{0i} + \left(\vQ\overline{\vX}\right)^{(j)}_i \Delta p_i $, 
employing $\vp_0$ and $\Delta \vp$ in Table~\ref{table:SLy5-min.Parameter}.
The correlated parameters $\vX^{(j)}$ satisfy the correlation $\mathcal{C}$ and its each input is $\sim \mathcal{N}(p_{0i}, \Delta p_i)$.
These are correlated random parameter sets of SLy5-min.

Using these randomized parameter sets $\vX^{(j)}$, we perform the RPA calculations $n$ times.
The RPA solver is \texttt{skyrme\_rpa} \cite{Colo13}.
The \texttt{skyrme\_rpa} solves the RPA equation for spherical nuclei in coordinate representation.
The calculation space is a sphere with radius 25 fm and mesh span 0.1 fm.
The dipole strength distribution is computed by smearing the resulting strength functions with a width $\Gamma=2$ MeV.
From this dipole strength distribution, we extract the peak energy of the GDR of $j$th random parameter, $E^{(j)}_\mathrm{GDR}$.
This calculated $E^{(j)}_\mathrm{GDR}$ is different from one that the original SLy5-min parameters $\vp_0$ produces 
because different parameter set $\vX^{(j)}$ is used.
These are repeated $n= 1000$ times for each random parameter set $\vX^{(j)}$. The mean value 
and the standard deviation of the calculated GDR peak energy are obtained by statistical processing from  
these 1000 results. 
This is the evaluation of uncertainty of $E_\mathrm{GDR}$ propagated from the uncertainties of the parameters.

\section{Results and discussion}
\label{sec:result}

\begin{figure}[!thb]
\begin{center}
\includegraphics[width=0.485\textwidth,keepaspectratio]{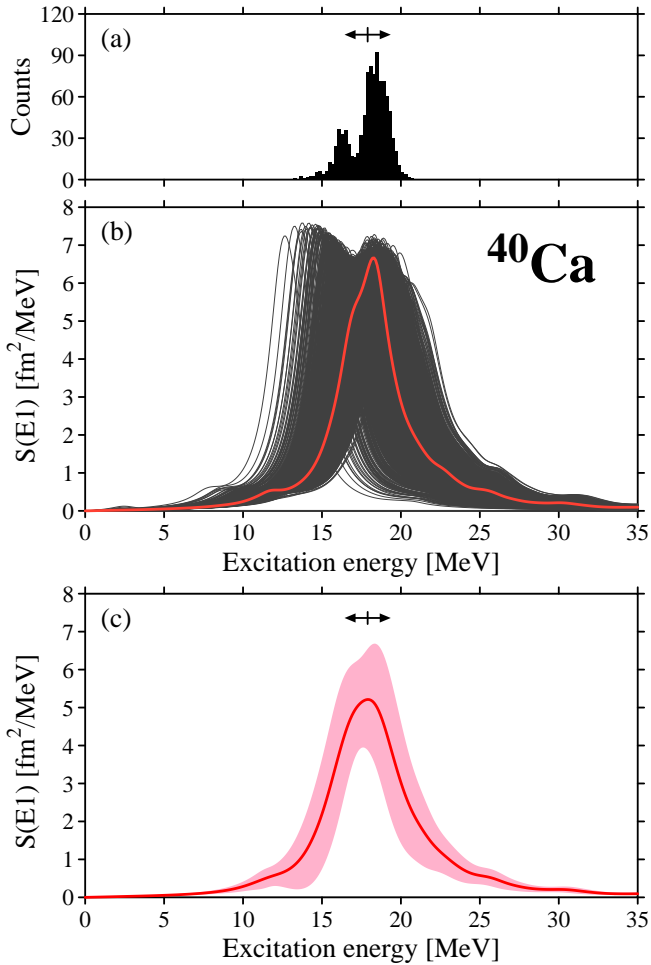}
\caption{$^{40}$Ca $E1$ strength distribution calculated with randomized parameter sets of SLy5-min interaction. (a) 
histogram of peak energies. 
(b) $E1$ strength distribution calculated with the original SLy5-min parameter (red) and those with randomized parameters (black). 
(c) Averaged $E1$ strength (red) with standard deviation (shaded red area).
Arrows in (a) and (c) denote the mean values of randomized peak energies and standard deviations.}
\label{40Ca.random.band}
\end{center}
\end{figure}

The dipole ($E1$) strength distributions $S(E1)$ in $^{40}$Ca calculated with the random parameter sets are plotted in Figs.~\ref{40Ca.random.band}(a)(b).
In Fig~\ref{40Ca.random.band}(b), the red line shows the $E1$ strength calculated with the original SLy5-min parameter set, 
and black lines are those with the random parameter sets.
Figure~\ref{40Ca.random.band}(a) shows histogram of peak energy of the randomized GDR.
Arrows denote the mean value of the peak energy and its standard deviation, $17.9\pm 1.5$ MeV.
Amazingly, the randomized GDR peak energies have two regions separated in its energy. 
In upper energy region, the randomized peak energies are scattered around $E_\mathrm{GDR}=18.3$ MeV that the original SLy5-min results.
In the lower energy region at $\lesim$ 17 MeV, the bunch is clearly separated from the upper one, though number of the randomized peak energies are small.
These peaks originate from a shoulder at 17 MeV of original $E1$ strength. [See red line in Fig.~\ref{40Ca.random.band}(b).]

Figure~\ref{40Ca.random.band}(c) shows averaged $E1$ strength distribution and the standard deviations. 
If the Skyrme parameter uncertainties are small, $|p_i - \Delta p_i |/ |p_i| \ll 1$, 
the averaged $E1$ strength is expected to be close to the original $E1$ strength calculated 
with the original SLy5-min parameter [the red line in Fig.~\ref{40Ca.random.band}(b)].
However, the averaged $E1$ strength is different from the original $E1$ strength.
The peak height is lowered and the GDR width is broadened.

\begin{figure*}[!thb]
\begin{center}
\includegraphics[width=0.900\textwidth,keepaspectratio]{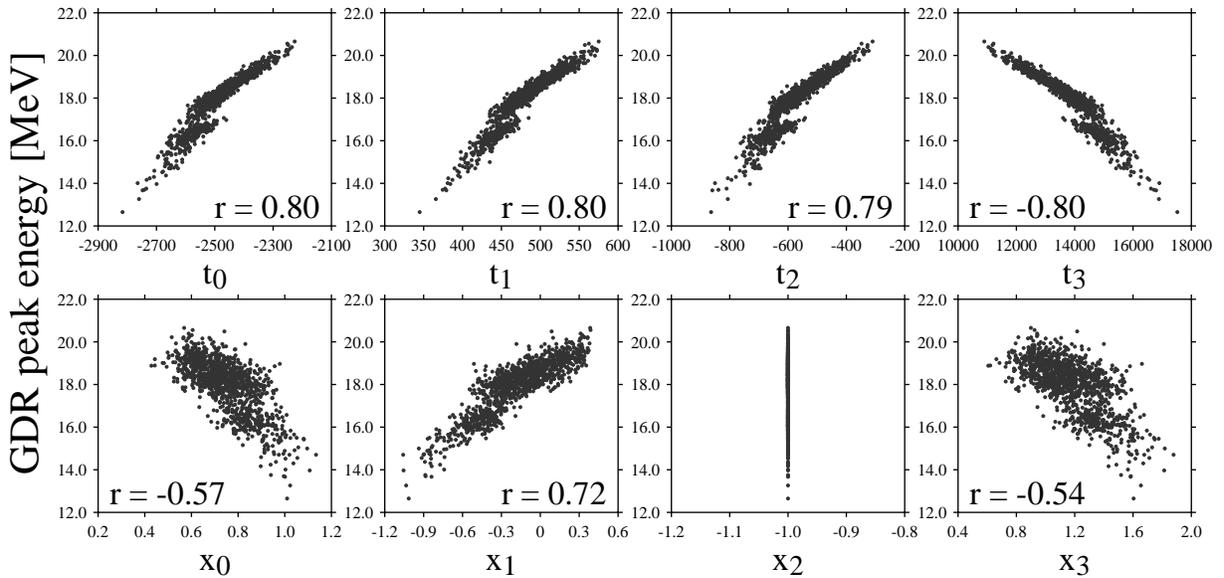}
\caption{Relations between the randomized GDR peak energies and Skyrme parameters $t_0$, $t_1$, $t_2$, $t_3$, $x_0$, $x_1$, $x_2$ and $x_3$, and Pearson correlation coefficients.}
\label{correlations}
\end{center}
\end{figure*}

To see which parameter affects the GDR peak energy, we calculated the Pearson correlation coefficients of 
the GDR peak energies and the Skyrme parameters $p_i$.
Figure~\ref{correlations} shows relations of the randomized Skyrme parameter and the corresponding randomized GDR energies.
Note that the parameters $x_2$, $W_0$, $W^\prime_0$, and $\sigma$ are fixed in SLy5-min parameter, and 
therefore the relations of $W_0$, $W^\prime_0$, and $\sigma$ are not plotted and always $x_2=-1$.
The randomized GDR peak energies have strong correlations with $t_0$, $t_1$, $t_2$, and $t_3$, 
and their Pearson correlation coefficients are $0.8$ or $-0.8$. 
Since the randomized GDR peak energies has two separated bunches, the plotted relations are also separated into upper- and lower-energy regions.
If we pick up the upper-energy region only, the Pearson correlation coefficients become larger.

Similar values of these Pearson correlation coefficients are related with strong correlations between $t_0$, $t_1$, $t_2$, and $t_3$,
shown in Table~\ref{table:correlationMatrix}. 
The moduluses of the correlations between $t_0$, $t_1$, $t_2$, and $t_3$ are larger than 0.95. This means 
that the uncertainties of $t_1$, $t_2$, and $t_3$ are almost redundant and can be approximated by the uncertainty of $t_0$.
Similar is seen in the relations of the randomized GDR peak energies and the parameters $x_0$, $x_1$, and $x_3$.
The moduluses of the correlations between $x_0$, $x_1$, and $x_3$ are larger than 0.93.
The strong correlation between the randomized GDR peak energies and the Skyrme parameters enable us to improve 
the calculated GDR peak energy by tuning the Skyrme parameters.

\begin{figure*}[!thb]
\begin{center}
\includegraphics[width=0.990\textwidth,keepaspectratio]{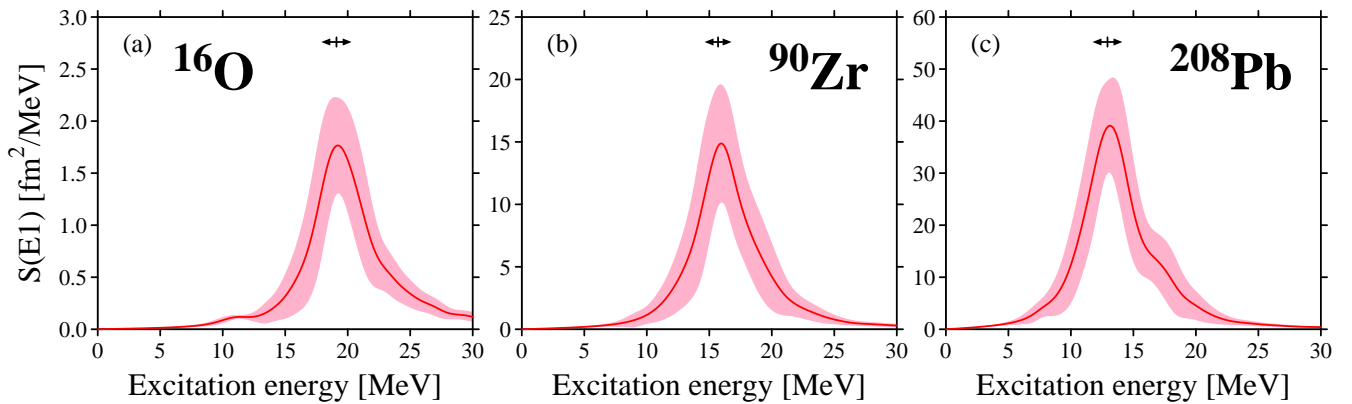}
\caption{Same as Fig.~\ref{40Ca.random.band}(c) but for (a) $^{16}$O, (b) $^{90}$Zr, and (c) $^{208}$Pb.}
\label{16O.90Zr.208Pb}
\end{center}
\end{figure*}

Same Monte Carlo calculations are performed for $^{16}$O, $^{90}$Zr, and $^{208}$Pb.
Similar to $^{40}$Ca [Fig.~\ref{40Ca.random.band}(b)], the randomized peak energies have two bunches in three calculated nuclei, 
but the mean energies of each bunch are close. 
Figures~\ref{16O.90Zr.208Pb} show the averaged $E1$ strength distributions and the standard deviations in $^{16}$O, $^{90}$Zr, and $^{208}$Pb.
The averaged peak energies and their standard derivations are listed in Table~\ref{table:EGDR}.
The standard deviations of the GDR peak energies are 1.0-1.5 MeV, irrespective of nuclear mass.
The Pearson correlation coefficients between the randomized GDR peak energies and the randomized Skyrme parameters $t_{0,1,2,3}$ 
in $^{16}$O, $^{90}$Zr and $^{208}$Pb are strong, similar to these in $^{40}$Ca. 
Those moduluses are approximately 0.83, 0.86, and 0.65, respectively. 
Also, the correlation coefficients of $x_{0,1,3}$ are almost same as those in $^{40}$Ca.
The uncertainty propagation from the Skyrme parameters to the GDR peak energies are not sensitive to the nuclear mass.

\begin{table}[!h]
\caption{Averaged peak energies and standard derivations of GDRs for $^{16}$O, $^{40}$Ca, $^{90}$Zr, and $^{208}$Pb, obtained by
Monte Carlo calculation using the randomized SLy5-min parameter sets.}
\begin{tabular}{c|rcl}
            &   average &     & deviation  \\ \hline
 $^{16}$O   &  19.10 & $\pm$ & 1.21 [MeV] \\
 $^{40}$Ca  &  17.89 & $\pm$ & 1.46 [MeV] \\
 $^{90}$Zr  &  15.70 & $\pm$ & 1.08 [MeV] \\
 $^{208}$Pb &  12.93 & $\pm$ & 1.19 [MeV] \\
\end{tabular}
\label{table:EGDR}
\end{table}

\begin{figure}[!thb]
\begin{center}
\includegraphics[width=0.485\textwidth,keepaspectratio]{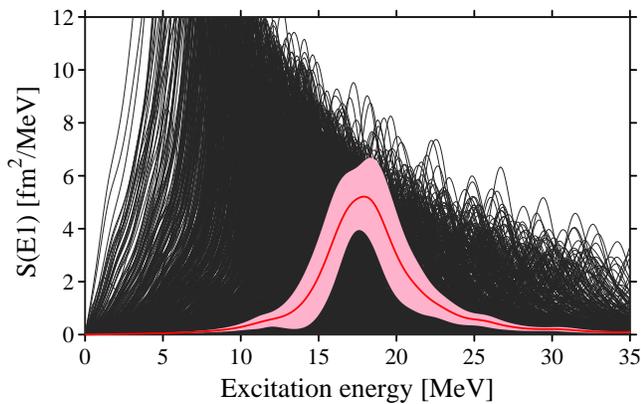}
\caption{$^{40}$Ca $E1$ strength distribution calculated using randomized uncorrelated SLy5-min parameter sets.
The red curve and shaded red area are same as Fig.~\ref{40Ca.random.band}(c).}
\label{40Ca.random.band.uncorrelated}
\end{center}
\end{figure}

Here, we show importance of the correlation matrix $\mathcal{C}$ for uncertainty evaluation.
We perform the same Monte Carlo calculation for $^{40}$Ca with neglecting $\mathcal{C}$, namely, 
we use the uncorrelated input $X^{(j)}_i = p_{0i} + \overline{\vX}^{(j)}_i \Delta p_i $ 
instead of the correlated input $X^{(j)}_i = p_{0i} + \left(\vQ\overline{\vX}\right)^{(j)}_i \Delta p_i$.
The calculated $E1$ strength distributions are plotted in Fig.~\ref{40Ca.random.band.uncorrelated} where 
the result with the correlated inputs [Fig.~\ref{40Ca.random.band}(c)] is also shown for comparison.
The randomized GDR peak energies scatter in wide energy region, 5-50 MeV.
The GDR peak height are inversed proportional to the excitation energy, attributed to the energy-weighted sum rule.
All the Pearson correlation coefficients between the GDR peak energies and the uncorrelated Skyrme parameters are small.
Those moduluses are less than 0.1.
Therefore the strong correlation between the GDR peak energy and Skyrme parameters comes from the correlations matrix $\mathcal{C}$.\\

The uncertainty of the GDR peak energy, needless to say, depends on the Skyrme parameter set and its uncertainties.
When we employ SAMi interaction \cite{SAMi} which has small uncertainties, $|p_i - \Delta p_i |/ |p_i| \ll 1$,
the randomized GDR peak energy has the uncertainty of $\sim$ 100 keV, even though the correlations between the parameters are not considered.

\section{Conclusion}
\label{sec:conclusion}

We performed the total Monte Carlo calculation to evaluate the uncertainty of the GDR peak energy, propagated from the uncertainties of the Skyrme parameters.
The RPA calculations with randomized Skyrme parameters is applied to spherical nuclei, $^{16}$O, $^{40}$Ca. $^{90}$Zr, and $^{208}$Pb.
In the case that SLy5-min parameter is employed with correlations between the parameters, 
the standard deviation of the GDR peak energy is $\sim 1$ MeV, irrespective of nuclear mass.
The GDR peak energy has strong correlations with the Skyrme parameters. 
These correlations enable us to handle the calculated GDR peak energy by tuning the Skyrme parameters. 
If we employ uncorrelated randomized Skyrme parameters, the GDR peak energy has no correlation with the Skyrme parameters.

The correlations between parameters of effective interactions are significant for accurate uncertainty evaluation.
We are evaluating uncertainties of other observables and will show them in a forthcoming paper.\

\section*{Acknowledgments}

The author thanks M. Kimura, Y. Utsuno, and S. Chiba for fruitful discussions and comments.


\begin{thebibliography}{99}


\bibitem{Stoitsov03} 
M.V. Stoitsov, J. Dobaczewski, W. Nazarewicz, S. Pittel, and D.J. Dean, 
Phys. Rev. C {\bf 68}, 054312 (2003).

\bibitem{Terasaki06} 
J. Terasaki and J. Engel,
Phys. Rev. C {\bf 74}, 044301 (2006).

\bibitem{Terasaki07} 
G. F. Bertsch, M. Girod, S. Hilaire, J.-P. Delaroche, H. Goutte, and S. P\'eru, 
Phys. Rev. Lett. {\bf 99}, 032502 (2007).

\bibitem{Hilaire07} 
S. Hilaire and M. Girod, 
Eur. Phys. J. A {\bf 33}, 237 (2007).

\bibitem{Terasaki08} 
J. Terasaki, J. Engel, and G. F. Bertsch, 
Phys. Rev. C {\bf 78}, 044311 (2008).

\bibitem{Inakura11} 
T. Inakura, T. Nakatsukasa, and K. Yabana, 
Phys. Rev. C {\bf 84}, 021302 (2011).

\bibitem{Ebata17} 
S. Ebata and T. Nakatsukasa, 
Phys. Scr. {\bf 92}, 064005 (2017).

\bibitem{Bulgac18} 
A. Bulgac, M. M. Forbes, S. Jin, R. N. Perez, and N. Schunck, 
Phys. Rev. C {\bf 97}, 044313 (2018).

\bibitem{Fattoyev11}
F.J. Fattoyev and J. Piekarewicz,
Phys. Rev. C {\bf 84}, 064302 (2011).

\bibitem{Gao13}
Y. Gao, J. Dobaczewski, M. Kortelainen, J. Toivanen, and D. Tarpanov,
Phys. Rev. C {\bf 87}, 034324 (2013).

\bibitem{Kortelainen13}
M. Kortelainen, J. Erler, W. Nazarewicz, N. Birge, Y. Gao, and E. Olsen,
Phys. Rev. C {\bf 88}, 031305(R) (2013).

\bibitem{Reinhard13}
P.-G. Reinhard, Piekarewicz, W. Nazarewicz, B.K. Agrawal, N. Paar, and X. Roca-Maza,
Phys. Rev. C {\bf 88}, 034325 (2013).

\bibitem{Goriely14}
S. Goriely and R. Capote,
Phys. Rev. C {\bf 89}, 054318 (2014).

\bibitem{Kortelainen15}
M. Kortelainen, 
J. Phys. G: Nucl. Part. Phys. {\bf 42}, 034021 (2015).

\bibitem{Erler15} 
J. Erler and and P.-G. Reinhard, 
J. Phys. G: Nucl. Part. Phys. {\bf 42}, 034026 (2015).

\bibitem{Roca-Maza15}
X. Roca-Maza, N. Paar, and G. Col\`o
J. Phys. G: Nucl. Part. Phys. {\bf 42}, 034033 (2015).

\bibitem{Reinhard16}
P.-G. Reinhard and W. Nazarewicz, 
Phys. Rev. C {\bf 93}, 051305(R) (2016).

\bibitem{Haverinen17} 
T. Haverinen and M. Kortelainen,
J. Phys. G: Nucl. Part. Phys. {\bf 44}, 044008 (2017).

\bibitem{Kejzlar20} 
V. Kejzlar, L. Neufcourt, W. Nazarewicz, and P.-G. Reinhard,
J. Phys. G: Nucl. Part. Phys. {\bf 47}, 094001 (2020).

\bibitem{Sprouse20}
T.M. Sprouse, R. Navarro Perez, R. Surman, M.R. Mumpower, G.C. McLaughlin, and N. Schunck,
Phys. Rev. C {\bf 101}, 055803(R) (2020).

\bibitem{Dobaczewski14} 
J. Dobaczewski, W. Nazarewicz, and P.-G. Reinhard, 
J. Phys. G {\bf 41}, 074001 (2014).

\bibitem{Brandt} 
S. Brandt, \textit{Statistical and Computational Methods in Data Analysis} (Springer, New York, 1997).

\bibitem{SLy5-min}
X. Roca-Maza, N. Paar, and G. Col\`o,
J. Phys. G {\bf 42}, 034033 (2015)

\bibitem{SLy5-min.correlation}
X. Roca-Maza, ``Covariance analysis and an example: SLy5-min," Materials of workshop on Information and Satistics in Nuclear Experiment and Theory, unpublished.
https://indico.cern.ch/event/253381/attachments/442168/613405/correlations.pdf

\bibitem{Colo13}.
C. Col\`o, L. Cao, N. V. Giai, and L. Capelli.
Comp. Phys. Commu. {\bf 184} (2013) 142-161.

\bibitem{SAMi}.
X. Roca-Maza, G. Col\`o, and H. Sagawa., Phys. Rev. C {\bf 86}, 031306(R) (2012).

\end{thebibliography}
\end{document}